\documentclass[a4paper,final]{ws-jnmp}

\begin{document}


\newcommand{\eref}[1]{(\ref{#1})}

\newcommand{\myShiftedP}[2]{\mathbb{E}_{#1}{#2}}
\newcommand{\myShiftedPi}[2]{\mathbb{E}_{#1}^{-1}{#2}}
\newcommand{\myShiftedN}[2]{\overline{\mathbb{E}}_{#1}{#2}}
\newcommand{\myShiftedNi}[2]{\overline{\mathbb{E}}_{#1}^{-1}{#2}}

\newcommand{\myshiftedP}[1]{\hat{#1}}
\newcommand{\myshiftedPi}[1]{\check{#1}}
\newcommand{\myshiftedN}[1]{\bar{#1}}
\newcommand{\myshiftedNi}[1]{\mathring{#1}}

\newcommand{\mymatrix}[1]{\mathsf{#1}}
\newcommand{\mysup}[1]{^{\scriptscriptstyle #1}} 
\newcommand{\mymu}{\beta_{0}}


\markboth{V.E. Vekslerchik}
{Functional representation of the negative AKNS hierarchy}

%
\catchline{19}{3}{2012}{}{}
%

\copyrightauthor{V.E. Vekslerchik}

\title{Functional representation of the negative AKNS hierarchy.}

\author{V.E. Vekslerchik}

\address{
Usikov Institute for Radiophysics and Electronics \\
12, Proskura st., Kharkov, 61085, Ukraine \\
\email{vekslerchik@yahoo.com}
}

\maketitle

\begin{history}
\received{29 January 2012}
\accepted{26 April 2012}
\end{history}

\begin{abstract}
This paper is devoted to the negative flows of the AKNS hierarchy.
The main result of this work is the functional representation of the extended 
AKNS hierarchy, composed of both positive (classical) and negative flows. 
We derive a finite set of functional equations, constructed by means of the 
Miwa's shifts, which contains all equations of the hierarchy.
Using the obtained functional representation we convert the nonlocal 
equations of the negative subhierarchy into local systems of higher order, 
derive the generating function of the conservation laws and the 
N-dark-soliton solutions for the extended AKNS hierarchy. 
As an additional result we obtain the  functional representation of the 
Landau-Lifshitz hierarchy.
\end{abstract}


\ccode{2010 Mathematics Subject Classification: 
  37J35, 
  35Q51, 
  35Q55, 
  37K10  
}

\section{Introduction. \label{sec-intro}}

This paper is devoted to the negative flows of the AKNS hierarchy.
The most well-known physical models described by this class of equations are the 
sine-Gordon and self-induced transparency models (see, \textit{e.g.}, books 
\cite{AS81,MB99} and references therein). 
Mathematically, the simplest way to describe the subject of this work is to use 
the zero-curvature representation (ZCR), the approach which is the 
base of the inverse scattering transform (IST). The AKNS hierarchy is an infinite 
set of equations coming as compatibility conditions for the system of linear 
equations consisting of the Zakharov-Shabat scattering problem \cite{ZS71,ZS73},
\begin{equation}
  \Psi_{x} = \mymatrix{U}(\lambda)\Psi,
  \qquad
  \mymatrix{U}(\lambda) = 
  i \begin{pmatrix} \lambda \; & R \cr Q \; & -\lambda \end{pmatrix}
\end{equation}
and a linear problem describing the evolution, 
\begin{equation}
  \Psi_{t} = \mymatrix{V}(\lambda)\Psi
\end{equation}
where $V(\lambda)$ is a matrix polynomial in $\lambda$.
The negative flows that form the negative AKNS hierarchy correspond to the 
case when $V(\lambda)$ is a polynomial in inverse powers of $\lambda$. 
Alternatively, negative (sub)hierarchies can be defined in terms of the 
recursion operator \cite{O77,FF81,V91,OR89}: if the hierarchy of higher-order 
symmetries (equations of the hierarchy) is generated 
by powers of the recursion operator, then the negative 
(sub)hierarchy can be constructed using its negative powers \cite{V91}. 
The recursion operator technique is widely exploited in the so-called 
``structural'' approach \cite{OR89} to integrable systems when one studies 
such questions as the Poisson structures, multi-Hamiltonian structures, 
mastersymmetries, $R$-matrices, \textit{etc}.

A characteristic feature of the negative flows is their nonlocality. 
Among large number of known negative hierarchies there are only
two, to our knowledge, examples of local ones: the Ablowitz-Ladik hierarchy, 
whose simlest positive and negative equations can be written as 
\begin{equation}
  i \frac{ \partial q_{n} }{ \partial t_{\pm 1} }
  = 
  \left( 1 + \kappa \left| q_{n} \right|^{2} \right) q_{n \pm 1},
 \qquad
 \kappa = \mbox{constant}
\end{equation}
and the Manna-Neveu generalization \cite{MN03} of the Hunter-Saxton equation 
\cite{HS91} that was discussed in \cite{B05}. Looking for the simplest 
``negative" matrices $V(\lambda)$, $V(\lambda) \propto \lambda^{-1}$, one 
arrives at \textit{nonlocal} equations. This nonlocality can be, in principle, 
eliminated by going to the equations that are not of the evolution type 
(the most well-known example is the sine-Gordon equation) or by introducing 
additional variables (as in equations that appear in the theory of the 
self-induced transparency). However, in any case the standard IST scheme is 
to be modified and probably because of this fact the negative AKNS hierarchy 
is not studied as comprehensively as the positive (classical) one.

The purpose of this paper, where we present some generalizations of the 
results obtained in \cite{AFGZ00,KP02,JZZ09}, is twofold. First, we want to 
describe the extended AKNS hierarchy composed of both positive and negative 
flows. To do this we will use an approach that can be viewed as an 
alternative to the ``traditional" ZCR or the method of \cite{AFGZ00}. 
The main result of this work is the functional representation of the positive 
and negative AKNS subhierarchies (Secs. \ref{sec-pos}-\ref{sec-neg}). We 
derive a finite set of functional equations, constructed by means of the 
Miwa's shifts, which contains all equations of the extended hierarchy, that 
can be recovered by the power series expansion. This approach has some 
advantages over the standard IST because in its framework one can avoid 
introducing the ``intermediate" objects of the inverse scattering method like 
Jost functions or scattering data and formulate results explicitly in terms 
of solutions (as, \textit{e.g.}, in the case of the generating function 
for the conservation laws presented in Sec. \ref{sec-pos}). 

The second goal of this paper is to obtain some explicit solutions, the 
dark soliton ones, of the extended AKNS system. Here we use another advandage 
of the functional representation which in many situations facilitates the 
calculations, especially when one deals with the whole hierarchy (AKNS 
hierarchy in our case) instead of one of its equations (say, the nonlinear 
Schr\"odinger equation (NLSE) in our case). As is shown in Sec. 
\ref{sec-dark}, we can enhance the classical results of \cite{ZS73} using 
simple algebraic proceeding.

The obtained results have some interesting byproducts. In Sec. \ref{sec-ll}, 
we derive the functional reprresentation of the Landau-Lifshitz hierarchy 
(LLH), that is known to be gauge equivalent to the AKNS hierarchy, and 
demonstrate that its negative part is symmetric to the positive one, which 
means that we have another example of local negative subhierarchy. Also we 
show that among the equations of the extended AKNS hierarchy one can find some 
models that were not associated with the AKNS system, such as, \textit{e.g.}, 
the Wadati-Konno-Ichikawa-like equations \cite{WKI1979} and Zakharov 
$(1+2)$-dimensional NLSE \cite{Z80}.

\section{Holonomy representation of the extended AKNS hierarchy. \label{sec-hr}}

Instead of the 
ZCR that is based on presenting the equations as the compatibility conditions 
for the linear systems
\begin{equation}
  \frac{\partial}{\partial t_{j}} \Psi = \mymatrix{U}_{j} \Psi, 
  \qquad
  \frac{\partial}{\partial \bar{t}_{k}} \Psi = \widetilde{\mymatrix{U}}_{k} \Psi, 
  \qquad
  j,k = 1, 2, ...
\end{equation}
we will be dealing with the linear systems constructed by the Miwa's shift 
operators that are applied to functions of a doubly infinite number of arguments,
\begin{equation}
  Q\left( \mathrm{t}, \bar{\mathrm{t}} \right) 
  = 
  Q\left( t_{1}, t_{2}, ... , \bar{t}_{1}, \bar{t}_{2}, ... \right) 
  = 
  Q\left( t_{j}, \bar{t}_{k} \right)_{j,k = 1, 2, ...}, 
\end{equation}
and are defined by
\begin{subequations}
\begin{equation}
    \mathbb{E}_{\xi} Q\left( \mathrm{t}, \bar{\mathrm{t}} \right) = 
    Q\left( \mathrm{t} + i[\xi], \bar{\mathrm{t}} \right),
\end{equation}
\begin{equation}
    \overline{\mathbb{E}}_{\eta} Q\left( \mathrm{t}, \bar{\mathrm{t}} \right) = 
    Q\left( \mathrm{t}, \bar{\mathrm{t}} + i [\eta] \right) 
\end{equation}
\end{subequations}
or 
\begin{subequations}
\begin{eqnarray}
&&
    \mathbb{E}_{\xi} 
    Q( t_{j}, \bar{t}_{k} )_{j,k = 1, 2, ...} 
    = 
    Q( t_{j} + i \xi^{j} / j, \bar{t}_{k} )_{j,k = 1, 2, ...}, 
\\[1mm]&&
    \overline{\mathbb{E}}_{\eta} 
    Q( t_{j}, \bar{t}_{k} )_{j,k = 1, 2, ...} 
    = 
    Q( t_{j}, \bar{t}_{k} + i \eta^{k} / k )_{j,k = 1, 2, ...}.
\end{eqnarray}
\end{subequations}
Thus, our goal is to study the compatibility conditions of the systems of the 
following type:
\begin{equation}
  \mathbb{E}_{\xi} \Psi = 
  \mymatrix{L}(\xi) \Psi
\label{holonomy-P}
\end{equation}
and
\begin{equation}
  \overline{\mathbb{E}}_{\eta} \Psi = 
  \bar{\mymatrix{L}}(\eta) \Psi,
\label{holonomy-N}
\end{equation}
where $\mymatrix{L}$ and $\bar{\mymatrix{L}}$ are $2 \times 2$-matrices,
that are given by 
\begin{subequations}
\label{commutativity}
\begin{eqnarray}
  \left[ \, \myShiftedP{ \xi_{1} }{} \mymatrix{L}\left(\xi_{2}\right) \, \right] 
  \mymatrix{L}\left(\xi_{1}\right) 
  & = & 
  \left[ \, \myShiftedP{ \xi_{2} }{} \mymatrix{L}\left(\xi_{1}\right) \, \right] 
  \mymatrix{L}\left(\xi_{2}\right), 
\label{commutativity-PP}
\\ 
  \left[ \, \myShiftedP{\xi}{} \bar{\mymatrix{L}}\left(\eta\right) \, \right] 
  \mymatrix{L}\left(\xi\right) 
  & = & 
  \left[ \, \myShiftedN{\eta}{} \mymatrix{L}\left(\xi\right) \, \right] 
  \bar{\mymatrix{L}}\left(\eta\right), 
\label{commutativity-PN}
\\ 
  \left[ \, 
    \myShiftedN{ \eta_{1} }{} \bar{\mymatrix{L}}\left(\eta_{2}\right) 
  \, \right] 
  \bar{\mymatrix{L}}\left(\eta_{1}\right) 
  & = & 
  \left[ \, 
    \myShiftedN{ \eta_{2} }{} \bar{\mymatrix{L}}\left(\eta_{1}\right) 
  \, \right] 
  \bar{\mymatrix{L}}\left(\eta_{2}\right). 
\label{commutativity-NN}
\end{eqnarray}
\end{subequations}
Using the standard strategy of the ZCR, that consists in introducing an 
auxiliary parameter, $\zeta$, and looking for the matrices $\mymatrix{L}$ 
and $\bar{\mymatrix{L}}$ with simplest dependence on $\zeta$, one can come 
to the matrices that lead to the AKNS hierarchy. It turns out that the matrices 
$\mymatrix{L}$ and $\bar{\mymatrix{L}}$ should be \textit{linear} functions of 
$\zeta^{-1}$ and $\zeta$. Indeed, expanding, for example, \eref{holonomy-P} in 
the power series in $\xi$ one obtains an infinite set of equations of the 
following structure:
\begin{equation}
  \left[ 
    \partial_{j} 
    + \mathcal{D}\left( \partial_{j-1}, ..., \partial_{1} \right) 
  \right]\Psi = 
  \mymatrix{L}_{j} \Psi,
  \qquad j=1,2,...
\end{equation}
where $\partial_{j} = \partial / \partial t_{j}$.
Expressing recursively the derivatives $\partial_{j-1}$, ..., $\partial_{2}$ 
one can rewrite the last equation as
\begin{equation}
  \partial_{j} \Psi = 
  \left[ 
  \mymatrix{L}_{j} 
  - \tilde{\mathcal{D}}\left( \partial_{1}^{j}, ..., \partial_{1} \right) 
  \right]\Psi, 
  \qquad j=1,2,...
\end{equation}
If $\mymatrix{L}_{1}$ is linear in $\zeta^{-1}$, then, 
after replacing the powers of $\partial_{1}$ by powers of $\mymatrix{L}_{1}$ one 
arrives at  
\begin{equation}
  \partial_{n} \Psi = 
  \mymatrix{U}_{j} \Psi, 
  \qquad j=1,2,...
\end{equation}
where $\mymatrix{U}_{j}$ is a $j$th order polynomial in $\zeta^{-1}$. 
In such a way, the standard structure of the ZCR of the AKNS 
hierarchy is reproduced, with $\zeta^{-1}$ playing the role of the spectral parameter.
Omitting the details of the calculations we present here the ``minimal" solution of 
\eref{commutativity}
\begin{subequations}
\label{def-L}
\begin{equation}
  \mymatrix{L}(\xi) = 
  \begin{pmatrix} 
    1 - \xi/\zeta + \xi^{2} \left(\myShiftedP\xi{Q}\right) R \quad & \xi R \cr
    \xi \myShiftedP\xi{Q} & 1 
  \end{pmatrix}
\label{def-L-P}
\end{equation}
and 
\begin{equation}
  \bar{\mymatrix{L}}(\eta) = 
  \mymatrix{1} 
  - \frac{ \zeta\eta }{ 1 + \left( \myShiftedN\eta{q} \right) r }
    \begin{pmatrix}
       1 & r \cr 
       \myShiftedN\eta{q} \quad & \left( \myShiftedN\eta{q} \right) r 
    \end{pmatrix}
\label{def-L-N}
\end{equation}
\end{subequations}
where $\mymatrix{1}$ is the unit matrix and the functions $Q$, $R$, $q$ and 
$r$ are subjected to some constraints that will be discussed below. 

In the following sections we study the systems of equations 
(the subhierarchies of the extended AKNS hierarchy) that appear as the result 
of \eref{commutativity} combined with \eref{def-L}.

\section{Positive AKNS subhierarchy. \label{sec-pos}}

The equations that follow from \eref{commutativity-PP} are 
\begin{equation}
  \left\{ 
  \begin{array}{l} 
  \left( \xi_{1} - \xi_{2} \right) 
    \myShiftedP{\xi_{1}}{} \myShiftedP{\xi_{2}}{} Q 
  = 
  \Lambda\left( \xi_{1},\xi_{2} \right) 
  \left( 
    \xi_{1} \, \myShiftedP{\xi_{1}}{Q}
    - 
    \xi_{2} \, \myShiftedP{\xi_{2}}{Q}
  \right), 
  \\[2mm]
  \left( \xi_{1} - \xi_{2} \right) R 
  = 
  \Lambda\left( \xi_{1},\xi_{2} \right) 
  \left( 
    \xi_{1} \, \myShiftedP{\xi_{2}}{R} 
    - 
    \xi_{2} \, \myShiftedP{\xi_{1}}{R} 
  \right) 
  \end{array}
  \right.
  \label{hie-PP}
\end{equation}
with 
\begin{equation}
  \Lambda\left(\xi_{1},\xi_{2}\right) 
  = 
  1 
  + \xi_{1}\xi_{2} 
    \left( \myShiftedP{\xi_{1}}{} \myShiftedP{\xi_{2}}{} Q \right) R. 
\label{def-Lambda}
\end{equation}
By simple algebra one can verify that \eref{hie-PP} ensure vanishing of all 
components of the matrix equation \eref{commutativity-PP}. 
The functional representation \eref{hie-PP} of the positive 
AKNS hierarchy has been derived in \cite{V02,PV02} (see also \cite{DM06b}).
It can be simplified by various limiting procedures. 
For example, sending $\xi_{2}$ to zero one arrives at the system that was used 
in \cite{PV02}:
\begin{equation}
  \left\{ 
  \begin{array}{rcl} 
  Q 
  - \myshiftedP{Q} 
  + i\xi \partial_{1}\myshiftedP{Q} 
  - \xi^{2} \myshiftedP{Q}^{2} R 
  & = & 0
  \\[2mm]
  \myshiftedP{R} - R - i\xi \partial_{1} R - \xi^{2} \myshiftedP{Q} R^{2} 
  & = & 0
  \end{array}
  \right.
\label{hie-akns-B}
\end{equation}
where the notation 
$\partial_{j} = \partial/\partial t_{j}$ and 
\begin{equation}
  \myshiftedP{Q} = \myShiftedP\xi{Q}  
\end{equation}
is used. 
Another reduction can be made by introducing the operator $\partial(\xi)$ by 
\begin{equation}
  \partial(\xi) = \sum_{j=1}^{\infty} \xi^{j} \partial_{j}. 
\end{equation}
Noting that
\begin{equation}
  \lim\limits_{ \xi_{1},\xi_{2} \to \xi } \; 
  \frac{ 1 }{ \xi_{1} - \xi_{2} } \; 
  \left( 
    \myShiftedP{\xi_{1}}{} \myShiftedPi{\xi_{2}}{} - 1 
  \right) f 
  = 
  i \xi^{-1} \partial(\xi) \, f 
\end{equation}
one can rewrite \eref{hie-PP} as 
\begin{equation}
  \left\{ 
  \begin{array}{l} 
  -i\partial(\xi) Q  = 
   Q - \Omega(\xi) \, \myshiftedP{Q} 
  \\[2mm]
  i\partial(\xi) R = 
  R - \Omega(\xi) \, \myshiftedPi{R}
  \end{array}
  \right.
\end{equation}
where 
\begin{equation}
  \Omega(\xi) = 
  \frac{ 1 }{ 1 + \xi^{2} \myshiftedP{Q} \myshiftedPi{R} } 
\label{def-Omega}
\end{equation}
and 
\begin{equation}
  \myshiftedPi{R} = \myShiftedPi\xi{R}.
\end{equation}

Expanding these functional equations in the power series in $\xi$ one obtains 
an infinite set of differential equations, the first non-trivial of which are the 
NLSE, 
\begin{equation}
  \left\{ 
  \begin{array}{rcl} 
  i\partial_{2} Q + \partial_{11} Q + 2 Q^{2}R 
  & = & 0
  \\ 
  -i\partial_{2} R + \partial_{11} R + 2 QR^{2} 
  & = & 0
  \end{array}
  \right.
\end{equation}
and the complex mKdV equation,
\begin{equation}
  \left\{ 
  \begin{array}{rcl} 
  \partial_{3} Q + \partial_{111} Q 
  + 6 QR \, \partial_{1} Q 
  & = & 0
  \\ 
  \partial_{3} R + \partial_{111} R 
  + 6 QR \, \partial_{1} R 
  & = & 0
  \end{array}
  \right.
\end{equation}
(here $\partial_{jk}$ stand for $\partial^{2} / \partial t_{j}\partial t_{k}$, \textit{etc})
that are the first equations of the AKNS hierarchy.

The positive AKNS subhierarchy, to repeat, is the classical AKNS hierarchy, 
that has been introduced in the early 1970's and which is one of the best-studied 
integrable systems. That is why we do not discuss Eq. \eref{hie-PP} here in details. 
The only thing that we need to illustrate some features of negative AKNS equations is the 
generating function for the conservation laws. 

\subsection*{Constants of motion.}

The AKNS hierarchy, as an an integrable system, possesses an infinite number 
of constants of motion that can be represented in the form  
\begin{equation}
  \mathcal{I}_{m}\left(t_{2},t_{3},...\right)
  = 
  \int_{\mathrm{reg}} I_{m}\left(t_{1}, t_{2},t_{3},...\right) dt_{1},
  \qquad 
  m=0,1,...
\end{equation}
(the symbol ``reg" indicates that one has to regularize, if necessary, the 
integrands by adding some constants depending on the boundary conditions that 
ensure the existence of the integrals).
As has been shown in \cite{PV02}, the generating function for $I_{m}$,
\begin{equation}
  I(\zeta) = \sum_{m=0}^{\infty} I_{m} \zeta^{m} 
\end{equation} 
has a very simple form when rewritten in terms of Miwa's shifts:
\begin{equation}
  I(\zeta) = \left( \myShiftedP\zeta{Q} \right) R. 
\end{equation} 
Indeed, Eqs. \eref{hie-PP} imply that 
\begin{equation}
  \partial(\xi) I(\zeta) 
  = 
  \partial_{1} J(\xi,\zeta)  
\label{akns-cl-E}
\end{equation} 
where 
\begin{equation}
  J(\xi,\zeta) 
  = 
  \xi\Omega(\xi) 
  \frac{ \bigl( \myShiftedP\xi{}\myShiftedP\zeta{} Q \bigr)
         \bigl( \myShiftedPi\xi{R} \bigr)}
       { 1 + \xi\zeta \bigl( \myShiftedP\xi{}\myShiftedP\zeta{} Q \bigr) R }   
\label{akns-cl-J}
\end{equation} 
which leads to $\partial_{j} \mathcal{I}_{m} = 0$ for any $j$ and $m$.
It should be noted that paper \cite{PV02} is devoted to the NLSE and not to 
the whole AKNS hierarchy, so a reader can find there only the particular 
case of Eqs. \eref{akns-cl-E} and \eref{akns-cl-J}. However, 
in order to not deviate 
from the main topic of this paper, we present them here ``as is'', leaving the 
proof for the a separate publication.

\section{Mixed AKNS subhierarchy. \label{sec-mix}}

The subhierarchy that is discussed in this section is closely related to the 
equations that usually appear in the works devoted to the negative flows of 
the AKNS hierarchy. The results presented below can be viewed as a 
generalization of the ones obtained in \cite{KP02,JZZ09}.
Substituting matrices \eref{def-L-P} and \eref{def-L-N} into 
\eref{commutativity-PN} one arrives at 
\begin{equation}
  \left\{
  \begin{array}{lcl} 
  \eta h(\eta) \myshiftedN{q} & = & \myshiftedN{Q} - Q, 
  \\[1mm]
  \eta h(\eta) r  & = & R - \myshiftedN{R} 
  \end{array}
  \right.
\label{hie-PN-A}
\end{equation}
and 
\begin{equation}
  \left\{
  \begin{array}{lcl}
  \myshiftedP{q} & = &  
  ( 1 - \xi \myshiftedP{q} R ) \left( q + \xi\myshiftedP{Q}\right) 
  \\ 
  r & = & 
  ( 1 - \xi \myshiftedP{Q} r ) \left( \myshiftedP{r} + \xi R \right) 
  \end{array}
  \right.
\label{hie-PN-B}
\end{equation}
where 
\begin{equation}
  h(\eta) = \frac{ 1 }{ 1 + \myshiftedN{q} r } 
\label{def-h}
\end{equation}
with the shortcuts that are used throughout this section:
\begin{equation}
  \myshiftedP{f} = \myShiftedP\xi{f}, 
  \qquad
  \myshiftedN{f} = \myShiftedN\eta{f}.  
\end{equation}
Thus, we have four equations for four functions $Q$, $R$, $q$ and $r$. 
Since we are discussing the AKNS hierarchy, our current task is to eliminate 
the last two and to obtain a closed system of equations for $Q$ and $R$.

Equations \eref{hie-PN-A} taken for two negative shifts, 
$\myShiftedN{\eta_{1}}{}$ and $\myShiftedN{\eta_{2}}{}$, can be rewritten 
in terms of the operator $\bar\partial(\eta)$,   
\begin{equation}
  \bar\partial(\eta) 
  = 
  \sum_{l=1}^{\infty} \eta^{l} \bar\partial_{l} 
  = 
  - i \eta  
  \lim\limits_{ \eta_{1},\eta_{2} \to \eta } \; 
  \frac{ 1 }{ \eta_{1} - \eta_{2} } \; 
  \left( 
    \myShiftedN{\eta_{1}}{} \myShiftedNi{\eta_{2}}{} - 1 
  \right) 
\end{equation}
(here $\bar\partial_{l} = \partial / \partial \bar{t}_{l}$) as 
\begin{subequations}
\begin{eqnarray}
  \phantom{-}i \eta^{-1} \bar\partial(\eta) \, Q 
  & = & 
  \omega\left( \eta \right) \myshiftedN{q}, 
\label{bp-eta-Q}
  \\ 
  - i \eta^{-1} \bar\partial(\eta) \, R 
  & = & 
  \omega\left( \eta \right) \myshiftedNi{r}, 
\label{bp-eta-R}
\end{eqnarray}
\end{subequations}
where
\begin{equation}
  \omega\left( \eta \right) 
  = 
  \frac{ 1 }{ 1 + \myshiftedN{q} \myshiftedNi{r} } 
\label{def-omega}
\end{equation}
and 
\begin{equation}
  \myshiftedNi{f} = \myShiftedNi\eta{f}. 
\end{equation}
Equations \eref{bp-eta-Q} and \eref{bp-eta-R} can be generalized 
by means of \eref{hie-PN-A} and \eref{hie-PN-B}, 
\begin{subequations}
\begin{eqnarray}
  \phantom{-}
  \left( 1 - \xi\eta \right) 
  i \eta^{-1} \bar\partial(\eta) \, \myshiftedP{Q} 
  & = & 
  \omega\left( \eta \right) 
  \left( 1 - \xi \myshiftedP{Q} \myshiftedNi{r} \right) 
  \left( \myshiftedN{q} + \xi \myshiftedP{Q} \right), 
  \\
  - \left( 1 - \xi\eta \right) 
  i \eta^{-1} \bar\partial(\eta) \, \myshiftedPi{R} 
  & = & 
  \omega\left( \eta \right) 
  \left( 1 - \xi \myshiftedN{q} \myshiftedPi{R} \right) 
  \left( \myshiftedNi{r} + \xi \myshiftedPi{R} \right) 
\end{eqnarray}
\end{subequations}
with
\begin{equation}
  \myshiftedPi{f} = \myShiftedPi\xi{f}. 
\end{equation}
Repeating this trick for two positive shifts, 
$\myShiftedP{\xi_{1}}{}$ and $\myShiftedP{\xi_{2}}{}$, one arrives at 
\begin{subequations}
\begin{eqnarray}
  \phantom{-}i \xi^{-1} \partial(\xi) \, \frac{ q }{ 1 + qr}  
  & = & 
  \frac{ \Omega(\xi) }{ 1 + qr} 
  \left( 1 - qr - 2\xi q \myshiftedPi{R} \right) \myshiftedP{Q}, 
  \\[2mm]
  - i \xi^{-1} \partial(\xi) \, \frac{ r }{ 1 + qr } 
  & = & 
  \frac{ \Omega(\xi) }{ 1 + qr }  
  \left( 1 - qr - 2\xi \myshiftedP{Q}r \right) \myshiftedPi{R}. 
\end{eqnarray}
\end{subequations}
These equations suffice for achieving our goal of eliminating $q$ and $r$. 
The resulting system can be written as 
\begin{equation}
  \left\{
  \begin{array}{lcl}
  0 & = & 
  \left( 1 - \xi\eta \right) \partial(\xi)\bar\partial(\eta) Q 
  + 2i\xi\beta(\eta) \partial(\xi) Q 
  + i \left[ \xi\eta - 2\xi\alpha(\xi) \right] \bar\partial(\eta) Q 
  + 2\xi\beta(\eta) Q, 
  \\[2mm]
  0 & = & 
  \left( 1 - \xi\eta \right) \partial(\xi)\bar\partial(\eta) R 
  - 2i\xi\beta(\eta) \partial(\xi) R 
  - i \left[ \xi\eta - 2\xi\alpha(\xi) \right] \bar\partial(\eta) R 
  + 2\xi\beta(\eta) R, 
  \end{array} 
  \right.
\label{hie-mix-A}
\end{equation} 
where the functions $\alpha$ and $\beta$ are defined by 
\begin{subequations}
\begin{eqnarray}
  \alpha(\xi) & = & 
  \frac{ \xi \myshiftedP{Q}\myshiftedPi{R} }
       { 1 + \xi^{2} \myshiftedP{Q}\myshiftedPi{R} }, 
  \\[2mm] 
  \beta(\eta) & = & 
  \frac{ \eta }{ 2 } \, 
  \frac{ 1 - \myshiftedN{q}\myshiftedNi{r} }
       { 1 + \myshiftedN{q}\myshiftedNi{r} } 
\end{eqnarray}
\end{subequations}
and are related by 
\begin{equation}
  \bar\partial(\eta) \alpha(\xi) = \partial(\xi) \beta(\eta). 
\label{hie-mix-Acl}
\end{equation}
The last identity, that can be verified directly, is very  important from the 
viewpoint of the conservation laws of the hierarchy.

Expanding Eqs. \eref{hie-mix-A} in the power series in $\xi$ and $\eta$ 
one arrives at 
\begin{equation}
  \left\{
  \begin{array}{lcl}
  0 & = & 
  \left( 
    \partial_{k+1}\bar\partial_{l+1} 
    - \partial_{k}\bar\partial_{l} 
    + 2i \beta_{l+1}\partial_{k}  
    - 2i \alpha_{k} \bar\partial_{l+1} 
  \right) Q, 
  \\[2mm]
  0 & = & 
  \left( 
    \partial_{k+1}\bar\partial_{l+1} 
    - \partial_{k}\bar\partial_{l} 
    - 2i \beta_{l+1}\partial_{k}  
    + 2i \alpha_{k} \bar\partial_{l+1} 
  \right) R, 
  \end{array} 
  \right.
  \quad
  k,l = 1,2,...
\label{hie-mix-C} 
\end{equation}
where
$\alpha_{k}$ and $\beta_{l}$ are the coefficient of the Taylor's expansion of 
the functions $\alpha(\xi)$ and $\beta(\eta)$, 
\begin{equation}
  \alpha(\xi) = \sum\limits_{k=1}^{\infty} \xi^{k} \alpha_{k}, 
  \qquad
  \beta(\eta) = \sum\limits_{l=1}^{\infty} \eta^{l} \beta_{l}.
\end{equation}

As in the case of positive subhierarchy, this two-parametric system can be 
simplified in several ways.  
First, Eqs. \eref{hie-mix-A} in the $\xi \to 0$ limit can be 
represented in the form 
\begin{equation}
  \left\{
  \begin{array}{lcl}
  0 & = & 
  \partial_{1}\bar\partial(\eta) Q 
  + i\eta\bar\partial(\eta) Q 
  + 2\beta(\eta) Q, 
  \\[2mm]
  0 & = & 
  \partial_{1}\bar\partial(\eta) R 
  - i\eta\bar\partial(\eta) R 
  + 2\beta(\eta) R, 
  \end{array} 
  \right.
\label{hie-mix-B}
\end{equation}
where $\beta(\eta)$ can be viewed as an additional dependent variable 
related to $Q$ and $R$ by 
\begin{equation}
  \partial_{1} \beta(\eta) = \bar\partial(\eta) QR 
\label{hie-mix-Bcl}
\end{equation}
which is the limiting form of \eref{hie-mix-Acl}.
Here one can see the nonlocality (explicit expression for $\beta(\eta)$ invokes 
$\partial_{1}^{-1}$ operator) that was observed in all works devoted to the 
negative AKNS flows. 
Equations \eref{hie-mix-B}
can be bilinearized by introducing the tau-functions $\tau$, $\sigma$ and $\rho$ by 
\begin{equation}
  \beta(\eta) = \partial_{1} \bar\partial(\eta) \ln\tau 
\end{equation}
and 
\begin{equation}
  Q = \frac{\sigma}{\tau}, 
  \qquad 
  R = \frac{\rho}{\tau}. 
\end{equation}
In new terms Eqs. \eref{hie-mix-B} and \eref{hie-mix-Bcl} become 
\begin{equation}
  \left\{
  \begin{array}{lcl}
  0 & = & 
  \left( D_{1} + i \eta \right) \bar{D}(\eta) \, \sigma \cdot \tau, 
  \\[2mm]
  0 & = & 
  \left( D_{1} - i \eta \right) \bar{D}(\eta) \, \rho \cdot \tau 
  \end{array} 
  \right.
\label{hie-mix-B-bilin}
\end{equation}  
and 
\begin{equation}
  D_{11} \tau \cdot \tau = 2 \rho\sigma, 
\label{hie-mix-tau}
\end{equation}
where $D_{1}$, $D_{11}$ and $\bar{D}(\eta)$ are the Hirota's bilinear operators,
\begin{equation}
  D_{j} \, u \cdot v = 
  \left( \partial_{j} u \right) v - u \left( \partial_{j} v \right) 
\end{equation}
$D_{jk...} = D_{j}D_{k}...$ etc., and 
\begin{equation}
  \bar{D}(\eta) = 
  \sum_{k=1}^{\infty} \eta^{k}\bar{D}_{k} 
\end{equation}
with
\begin{equation}
  \bar{D}_{k} \, u \cdot v = 
  \left( \bar\partial_{k} u \right) v - u \left( \bar\partial_{k} v \right). 
\end{equation}

Returning from the power series to ``individual" flows one can rewrite 
\eref{hie-mix-B} and \eref{hie-mix-B-bilin} as 
\begin{equation}
  \left\{
  \begin{array}{l}
  0 = 
  \left( 
    \partial_{1}\bar\partial_{l+1} + i\bar\partial_{l}  + 2\beta_{l+1} 
  \right) Q, 
  \\[2mm]
  0 = 
  \left( 
    \partial_{1}\bar\partial_{l+1} - i\bar\partial_{l}  + 2\beta_{l+1} 
  \right) R, 
  \\[2mm] 
  \partial_{1} \beta_{l} = \bar\partial_{l} QR, 
  \end{array} 
  \right.
  \hspace{20mm}
  l = 1,2,...
\label{hie-mix-D} 
\end{equation}
and 
\begin{equation}
  \left\{
  \begin{array}{lcl}
  0 & = & 
  \left( D_{1} \bar{D}_{l+1} + i \bar{D}_{l} \right) \sigma \cdot \tau, 
  \\[2mm]
  0 & = & 
  \left( D_{1} \bar{D}_{l+1} - i \bar{D}_{l} \right) \rho \cdot \tau, 
  \end{array} 
  \right.
  \hspace{20mm}
  l = 1,2,...
\label{hie-mix-D-bilin} 
\end{equation}
together with \eref{hie-mix-tau}. 

Equations \eref{hie-mix-D}  and \eref{hie-mix-D-bilin} were discussed in 
\cite{KP02,JZZ09}, thus Eqs. \eref{hie-mix-B} and \eref{hie-mix-B-bilin} 
can be viewed as their compact form while Eqs. \eref{hie-mix-A} and 
\eref{hie-mix-C} as their generalization.

The simplest (and hence most representative) equation of the mixed AKNS 
subhierarchy, discussed in this section, can be written as 
\begin{equation}
  \left\{
  \begin{array}{lcl}
  0 & = & Q_{xy} + 2PQ, 
  \\[2mm]
  0 & = & R_{xy} + 2PR, 
  \\[2mm] 
  P_{x} 
  & = & 
  \left( QR \right)_{y}, 
  \end{array} 
  \right.
\end{equation}
where $x = t_{1}$, $y = \bar{t}_{1}$ and $P = \beta_{1}$.
The next one (Eq. \eref{hie-mix-D} with $l=1$), 
\begin{equation}
  \left\{
  \begin{array}{rcl}
  - i Q_{t} & = &  Q_{xy} + 2PQ, 
  \\[2mm]
  i R_{t} & = & R_{xy} + 2PR, 
  \\[2mm] 
  P_{x} 
  & = & 
  \left( QR \right)_{y} 
  \end{array} 
  \right.
\end{equation}
($x = t_{1}$, $y = \bar{t}_{2}$, $t = \bar{t}_{1}$ and $P = \beta_{2}$) 
is nothing but the $(1+2)$-dimensional NLSE introduced by Zakharov \cite{Z80}.

\subsection*{Constants of motion.}

In Sec. \ref{sec-pos}, we have presented the generating function for the 
constants of motion, $I(\zeta) = \left( \myShiftedP\zeta{Q} \right) R$, for 
the positive (classical) AKNS hierarchy. It turns out that equation similar to 
\eref{akns-cl-E} holds for the negative flows as well: 
\begin{equation}
  \bar\partial(\eta) I(\zeta) 
  = 
  \partial_{1} \bar{J}(\eta,\zeta), 
\label{akns-cl-EN}
\end{equation} 
where 
\begin{equation}
  \bar{J}(\eta,\zeta) 
  = 
  \frac{ \eta }{ 1 - \zeta\eta } \, 
  \omega(\eta) \left[ 
    1 - \zeta \left( \myShiftedP\zeta{Q} \right) \bigl( \myShiftedNi\eta{r} \bigr) 
  \right].
\label{akns-cl-JN}
\end{equation} 
This means that the quantities $I_{m}$, as is expected, are the constants of 
both positive and negative AKNS subhierarchies.
Again, the proof of \eref{akns-cl-EN} and \eref{akns-cl-JN} will be published 
elsewhere.

\section{Negative AKNS subhierarchy. \label{sec-neg}}

Equations that follow from the commutativity condition \eref{commutativity-NN} 
can be written as 
\begin{equation}
  \left\{
  \begin{array}{lcl}
  \eta_{1}^{-1} \left( \myShiftedN{\eta_{1}}{} - 1 \right) 
  h( \eta_{2} ) \myShiftedN{\eta_{2}}{q}
  & = & 
  \eta_{2}^{-1} \left( \myShiftedN{\eta_{2}}{} - 1 \right) 
  h( \eta_{1} ) \myShiftedN{\eta_{1}}{q}, 
  \\[2mm]
  \eta_{1}^{-1} \left( \myShiftedN{\eta_{1}}{} - 1 \right) 
  h( \eta_{2} ) r 
  & = & 
  \eta_{2}^{-1} \left( \myShiftedN{\eta_{2}}{} - 1 \right) 
  h( \eta_{1} ) r, 
  \end{array}
  \right.
\label{eq-neg-qr}
\end{equation}
where $h(\eta)$ is defined by \eref{def-h}. This is a closed system for the 
functions $q$ and $r$ which is closely related to the LLH 
that is discussed in Sec. \ref{sec-ll}. 
However, as in Sec. \ref{sec-mix}, we use the words 
``negative AKNS (sub)hierarchy" bearing in mind another system, the one for 
the functions $Q$ and $R$ that can be written in terms of Miwa's shifts and 
differential operators with respect to negative ``times", 
$\myShiftedN{\eta}{}$ and $\bar\partial_{j} = \partial / \partial\bar{t}_{j}$.

There are several ways to eliminate $q$ and $r$ together with 
$\myShiftedP{\xi}{}$ from Eqs. \eref{commutativity-PN} and 
\eref{commutativity-NN}. The shortest one can be described as follows. 
Passing from $\myShiftedN{\eta}{}$ to $\bar\partial(\eta)$ one can rewrite 
\eref{eq-neg-qr} as 
\begin{equation}
  \left\{ 
  \begin{array}{rcl}
  i \bar\partial\left( \eta \right) \omega_{0} q 
  & = & 
  \kappa \omega(\eta) \myshiftedN{q} - \gamma(\eta) \omega_{0} q, 
  \\[2mm]
  - i \bar\partial\left( \eta \right) \omega_{0} r 
  & = & 
  \kappa \omega(\eta) \myshiftedNi{r} - \gamma(\eta) \omega_{0} r, 
  \end{array} 
  \right. 
\end{equation}
where $\omega(\eta)$ is defined in \eref{def-omega},
$\myshiftedN{q} = \myShiftedN\eta{q}$, 
$\myshiftedNi{r} = \myShiftedNi\eta{r}$, 
\begin{equation}
  \gamma(\eta) = \frac{ 1 - \myshiftedN{q}\myshiftedNi{r} }
                 { 1 + \myshiftedN{q}\myshiftedNi{r} } 
  = 
  \frac{ 2 }{ \eta } \, \beta(\eta), 
\end{equation}
$\kappa = \gamma(0) = (1-qr)/(1+qr)$ and 
$\omega_{0} = \omega(0) = 1/(1+qr)$.
Combining the above formulae with Eqs. \eref{hie-PN-A} one can obtain the 
system
\begin{equation}
  \left\{
  \begin{array}{lcl}
  0 & = & 
  \bar\partial_{1} \bar\partial(\eta) Q 
  + i \eta^{-1} \kappa \bar\partial(\eta) Q 
  - i \gamma(\eta) \bar\partial_{1} Q, 
  \\[2mm]
  0 & = & 
  \bar\partial_{1} \bar\partial(\eta) R 
  - i \eta^{-1} \kappa \bar\partial(\eta) R 
  + i \gamma(\eta) \bar\partial_{1} R. 
  \end{array} 
  \right.
\label{eq-neg-QR}
\end{equation}
To finish the derivation of the negative AKNS equations one has to express 
$\gamma(\eta)$ in terms of $Q$ and $R$. This can be easily achieved by means 
of Eqs. \eref{hie-PN-A}:
\begin{equation}
  \gamma^{2}(\eta) = 
  1 
  - 4 \eta^{-2} 
    \left[ \bar\partial(\eta) Q \right] 
    \left[ \bar\partial(\eta) R \right]. 
\label{eq-neg-k}
\end{equation}
So we have a closed system of Eqs. \eref{eq-neg-QR} and \eref{eq-neg-k} 
that can be viewed as the functional representation of the negative AKNS 
subhierarchy.
Expanding \eref{eq-neg-QR} and \eref{eq-neg-k} in the power series in $\eta$ 
one arrives at an infinite number of partial differential equations describing 
the negative flows. The simplest one can be written as 
\begin{equation}
  \left\{
  \begin{array}{lcl}
  0 & = & 
  \left( 
    i \bar\partial_{2}  
    + \kappa^{2} \bar\partial_{1} \kappa^{-1} \bar\partial_{1} 
  \right) Q, 
  \\[2mm]
  0 & = & 
  \left( 
    i \bar\partial_{2}  
    - \kappa^{2} \bar\partial_{1} \kappa^{-1} \bar\partial_{1} 
  \right) R 
  \end{array} 
  \right.
\end{equation}
with 
\begin{equation}
  \kappa = 
  \sqrt{ 
    1 - 4 \left( \bar\partial_{1} Q \right) \left( \bar\partial_{1} R \right) 
  }.
\end{equation}
The second negative AKNS equation can be written as 
\begin{equation}
  \left\{
  \begin{array}{lcl}
  0 & = & 
  \left( 
      2\bar\partial_{3} 
    - \bar\partial_{111} 
    - 3i\kappa^{2}\bar\partial_{2}\kappa^{-1}\bar\partial_{1} 
  \right) Q, 
  \\[2mm]
  0 & = & 
  \left( 
      2\bar\partial_{3} 
    - \bar\partial_{111} 
    + 3i\kappa^{2}\bar\partial_{2}\kappa^{-1}\bar\partial_{1} 
  \right) R.
  \end{array} 
  \right.
\end{equation}
One can see that the negative AKNS equations have the same nonlinearity structure 
as the Wadati-Konno-Ichikawa equations \cite{WKI1979}, that have 
been derived as a generalization of the \textit{positive} AKNS hierarchy.

\section{Landau-Lifshitz hierarchy. \label{sec-ll}}

In this section we discuss the ``pure negative" equations, the ones stemming from  
\eref{commutativity-NN}, from a standpoint different from that of 
Sec. \ref{sec-neg}. To this end it is convenient to introduce the matrix 
\begin{equation}
  \bar{\mymatrix{S}}(\eta) = 
  \omega(\eta) 
  \begin{pmatrix}
    1 - \myshiftedN{q} \myshiftedNi{r} & 
    2 \myshiftedNi{r} \cr 
    2 \myshiftedN{q} & 
    -1 + \myshiftedN{q}\myshiftedNi{r} 
  \end{pmatrix},
\label{def-bar-S}
\end{equation}
where $\myshiftedN{q} = \myShiftedN\eta{q}$ and 
$\myshiftedNi{r} = \myShiftedNi\eta{r}$.
In principle, one can express $\bar{\mymatrix{S}}$ in terms of $Q$ and $R$ 
using~\eref{eq-neg-qr}, 
\begin{equation}
  \bar{\mymatrix{S}}(\eta) = 
  \begin{pmatrix}
    \gamma(\eta) & 
    -2i \eta^{-1} \bar\partial(\eta) R \cr 
     2i \eta^{-1} \bar\partial(\eta) Q & 
    - \gamma(\eta) 
  \end{pmatrix}.
\end{equation}
However, we will not use this relationship below, restricting ourselves to the 
consequences of \eref{commutativity-NN} that can be formulated in the terms of 
the matrix \eref{def-bar-S}. 
By straightforward algebra one can show that equations \eref{eq-neg-qr} lead 
to the following equations: 
\begin{equation}
  \eta_{1}^{-1} \bar\partial(\eta_{1}) \; 
  \bar{\mymatrix{S}}\left(\eta_{2}\right) 
  = 
  \eta_{2}^{-1} \bar\partial(\eta_{2}) \; 
  \bar{\mymatrix{S}}\left(\eta_{1}\right) 
\label{hie-LL-A}
\end{equation}
and 
\begin{equation}
  2i \eta \, \bar\partial_{1} \bar{\mymatrix{S}}(\eta) = 
  \left[ \, \bar{\mymatrix{S}}(\eta), \mymatrix{S} \, \right]\!, 
\label{hie-LL-B}
\end{equation}
where 
\begin{equation}
  \mymatrix{S} = \bar{\mymatrix{S}}(0). 
\end{equation}
Using \eref{hie-LL-A} one can rewrite \eref{hie-LL-B} as 
\begin{equation}
  2i \, \bar\partial(\eta) \mymatrix{S} = 
  \left[ \, \bar{\mymatrix{S}}(\eta), \mymatrix{S} \, \right]
\label{hie-LL-C}
\end{equation}
and calculate $\bar{\mymatrix{S}}(\eta)$ in terms of $\mymatrix{S}$: 
\begin{equation}
  \bar{\mymatrix{S}}(\eta) = 
  - i \mymatrix{S} \, \bar\partial(\eta) \mymatrix{S} 
  + \lambda(\eta) \, \mymatrix{S}. 
\label{hie-LL-CB}
\end{equation}
Here the function $\lambda(\eta)$ should be determined from the condition 
\begin{equation}
  \bar{\mymatrix{S}}^{2}(\eta) = \mymatrix{1}
\end{equation}
which leads to 
\begin{equation}
  \lambda^{2}(\eta) = 
  - \frac{1}{2} \mathop{\mbox{tr}} 
    \left[ \bar\partial(\eta) \mymatrix{S} \right] ^{2}.
\label{hie-LL-CC}
\end{equation}
Thus, we have a closed system of Eqs. \eref{hie-LL-C}, \eref{hie-LL-CB} 
and \eref{hie-LL-CC} for the matrix $\mymatrix{S}$.

Expanding these equations in the power series in $\eta$,
one can obtain an infinite set of equations 
\begin{eqnarray}
  0 & = & 
  i \bar\partial_{2} \mymatrix{S} 
  + \frac{1}{2} \left[ \, \mymatrix{S}'', \mymatrix{S} \right]\!, 
  \\[2mm] 
  0 & = & 
  \bar\partial_{3} \mymatrix{S} 
  + \mymatrix{S}''' 
  + \frac{3}{2} 
    \bigl( \left( \mymatrix{S}' \right)^{2} \mymatrix{S} \bigr)', 
  \\[2mm] 
   & ... & 
\nonumber
\end{eqnarray} 
where the symbol $'$ is used to denote the derivative with respect to 
$\bar{t}_{1}$:
\begin{equation}
  \mymatrix{S}' = \bar\partial_{1} \mymatrix{S}. 
\end{equation}
The above equations are the simplest equations of the LLH. 
In other words, we have shown that Eq. \eref{commutativity-NN} lead to the LLH 
and derived the functional representation of the latter. 

The fact that the LLH is closely related to the AKNS hierarchy is not new, it is 
known since the works of Zakharov and Takhtadzhyan \cite{ZT79}. 
However, we would like to note that the Landau-Lifshitz equation that was 
mentioned in \cite{KP02}, whose results are generalized in this section, 
and the Landau-Lifshitz equation that appear in \cite{ZT79} are not the 
same: the last one belongs to the positive subhierarchy, while the former 
describes the negative flows. 
This indicates that the symmetry between the positive and negative flows of the 
AKNS hierarchy, which is not visible in terms of $Q$ and $R$, becomes apparent
at the level of the LLH.

\section{Dark solitons of the extended AKNS hierarchy. \label{sec-dark}}

In this section we would like to present the dark-soliton solutions of the 
extended (describing both positive and negative flows) AKNS hierarchy. We will 
not derive them from scratch but will use the classical results (for the 
positive subhierarchy) and extend them to cover the negative flows. 
The $N$-soliton solutions for the NLSE were obtained in the beginning of the 
seventies by Zakharov and Shabat who developed in \cite{ZS71,ZS73} the 
corresponding version of the IST. Since all equations of the AKNS hierarchy, 
considered from the viewpoint of the inverse scattering approach, are based on 
the same scattering problem, their solutions (dark solitons in our case) 
possess the same structure that the ones derived in \cite{ZS73}.
Thus, to solve any equation of the hierarchy one can utilize a big part of the 
results of \cite{ZS73}. The only thing that one has to do is to establish some 
relations between parameters of the solutions (the so-called ``dispersion 
laws") which are different for different equations of the hierarchy. These 
considerations suggest the following procedure: we look for the solutions 
whose structure is similar to the classical dark solitons and then find, 
using some simple algebraic calculations, the conditions that convert them into 
solutions of all equations of the hierarchy (both positive and negative).

The main building blocks for the dark-soliton solutions of the AKNS hierarchy 
are $N \times N$ matrices $\mymatrix{A}$ that satisfy the ``almost rank-one" 
condition 
\begin{equation}
  \mymatrix{L} \mymatrix{A} 
  - 
  \mymatrix{A} \mymatrix{L}^{-1} 
  = 
  | \,\ell\, \rangle \langle a | 
\end{equation}
where $\mymatrix{L}$ is a constant diagonal matrix, $| \,\ell\, \rangle$ is 
constant $N$-component column, 
$| \,\ell\, \rangle = \left( \ell_{1}, ... , \ell_{N} \right)^{T}$
and $\langle a |$ is $N$-component row depending on the coordinates 
describing the AKNS flows,
$\langle a\left( \mathrm{t}, \bar{\mathrm{t}} \right) | = 
\left( 
  a_{1}\left( \mathrm{t}, \bar{\mathrm{t}} \right), ... , 
  a_{N}\left( \mathrm{t}, \bar{\mathrm{t}} \right) 
\right)$, 
and matrices $\mymatrix{H}_{\zeta}$ are defined by 
\begin{equation}
  \mymatrix{H}_{\zeta} 
  = 
  \left( \zeta \mymatrix{1} - \mymatrix{L} \right) 
  \left( \zeta \mymatrix{1} - \mymatrix{L}^{-1} \right)^{-1}, 
\end{equation}
where $\mymatrix{1}$ is the $N \times N$ unit matrix.

The remarkable property of the above matrices, that will be repeatedly used 
below, is that the determinants
\begin{equation}
  \omega\left( \mymatrix{A} \right) 
  = 
  \det \left| \mymatrix{1} + \mymatrix{A} \right|
\label{omega-def}
\end{equation}
satisfy the Fay-like identity
\begin{equation} 
  (\xi   - \eta)  \, \omega_{\zeta} \, \omega_{\xi\eta} 
  + 
  (\eta  - \zeta) \, \omega_{\xi}   \, \omega_{\eta\zeta} 
  +  
  (\zeta - \xi)   \, \omega_{\eta}  \, \omega_{\zeta\xi} 
  = 0, 
\label{fay}
\end{equation}
where
\begin{equation}
  \omega = \omega\left( \mymatrix{A} \right),
  \qquad
  \omega_{\zeta} = \omega\left( \mymatrix{A} \mymatrix{H}_{\zeta} \right),
  \qquad
  \omega_{\xi\eta} = 
  \omega\left( \mymatrix{A} \mymatrix{H}_{\xi} \mymatrix{H}_{\eta} \right).
\end{equation}
One can find an elementary proof of this identity in \ref{app-Fay}. 
It is shown below that upon representing the Miwa's shifts as multiplication 
by combinations of matrices $\mymatrix{H}_{\zeta}$ it is possible to derive from 
(\ref{fay}) the identities similar to equations (\ref{hie-PP}), 
(\ref{hie-PN-A}) and (\ref{eq-neg-qr}) that we want to solve.

\subsection{Solution of the equations of the positive subhierarchy.}

First let us study the positive subhierarchy. The key feature is to assume that 
the dependence on the positive ``times" $t_{1}, t_{2}, ... $ is governed by 
\begin{equation}
  \mathbb{E}_{\xi} \, \mymatrix{A} = 
  \mymatrix{A} \, \mymatrix{H}_{\alpha} \mymatrix{H}_{0}^{-1}, 
\end{equation}
where the function $\alpha = \alpha(\xi)$, $\alpha(0) = 0$, is specified below. 
Writing down Eq. (\ref{fay}) with 
$\xi=\alpha_{1}$, $\eta=\alpha_{2}$ and $\zeta=0$ and the matrix $\mymatrix{A}$ 
being replaced with $\mymatrix{A} \mymatrix{H}_{0}^{-1}$ one arrives at 
\begin{equation}
  \left( \alpha_{1} - \alpha_{2} \right) 
  \omega(\mymatrix{A})  
  \left[ \mathbb{E}_{1}\mathbb{E}_{2} \omega(\mymatrix{B}) \right]
  = 
  \alpha_{1} 
  \left[ \mathbb{E}_{1} \omega(\mymatrix{B}) \right] 
  \left[ \mathbb{E}_{2} \omega(\mymatrix{A}) \right] 
  - \alpha_{2} 
  \left[ \mathbb{E}_{1} \omega(\mymatrix{A}) \right] 
  \left[ \mathbb{E}_{2} \omega(\mymatrix{B}) \right]\!,
\label{fay-PB}
\end{equation}
where $\mymatrix{B} = \mymatrix{A} \mymatrix{H}_{0}$ and 
\begin{equation}
  \mathbb{E}_{k} = \mathbb{E}_{\xi_{k}}, 
  \qquad
  \alpha_{k} = \alpha\left( \xi_{k} \right)
  \qquad
  (k=1,2).
\end{equation}
In a similar way, Eq. (\ref{fay}) with 
$\xi=1/\alpha_{1}$, $\eta=\alpha_{2}$, $\zeta=0$ and 
$\mymatrix{A} \to \mymatrix{A} \mymatrix{H}_{0}^{-1}$ 
leads to 
\begin{equation} 
  \left( 1 - \alpha_{1}\alpha_{2} \right) 
  \omega(\mymatrix{A})  
  \left[ \mathbb{E}_{1}^{-1}\mathbb{E}_{2} \omega(\mymatrix{A}) \right]
  =  
  \left[ \mathbb{E}_{1}^{-1} \omega(\mymatrix{A}) \right] 
  \left[ \mathbb{E}_{2} \omega(\mymatrix{A}) \right] 
  - 
  \alpha_{1}\alpha_{2} 
  \left[ \mathbb{E}_{1}^{-1} \omega(\mymatrix{C}) \right] 
  \left[ \mathbb{E}_{2} \omega(\mymatrix{B}) \right]\!, 
\label{fay-PA}
\end{equation}
where $\mymatrix{C} = \mymatrix{A} \mymatrix{H}_{0}^{-1}$. 
Rewriting (\ref{fay-PB}) and (\ref{fay-PA}) in terms of functions $Q$ and $R$ 
defined by 
\begin{equation}
	Q = U \; \frac{ \omega(\mymatrix{B}) }{ \omega(\mymatrix{A}) }, 
	\qquad 
	R = V \; \frac{ \omega(\mymatrix{C}) }{ \omega(\mymatrix{A}) }, 
\label{def-QR}
\end{equation}
where $U$ and $V$ are two auxiliary functions one can obtain 
\begin{equation} 
  \mathbb{E}_{12}\, Q 
  = 
  \frac{ \left( \mathbb{E}_{12}\, U \right) }
       { \alpha_{1} - \alpha_{2} } 
  \;  
  \frac{ 
    \left[ \mathbb{E}_{1}\, \omega(\mymatrix{A}) \right] 
    \left[ \mathbb{E}_{2}\, \omega(\mymatrix{A}) \right] 
    }{
    \omega(\mymatrix{A}) \left[ \mathbb{E}_{12}\, \omega(\mymatrix{A}) \right] 
    } 
  \left[
    \frac{ \alpha_{1} }{ \mathbb{E}_{1}U } 
    \left( \mathbb{E}_{1}\, Q \right) 
    - 
    \frac{ \alpha_{2} }{ \mathbb{E}_{2}U  } 
    \left( \mathbb{E}_{2}\, Q \right) 
  \right]  
\end{equation}
and 
\begin{equation} 
  \frac{ 
    \left[ \mathbb{E}_{1}\, \omega(\mymatrix{A}) \right] 
    \left[ \mathbb{E}_{2}\, \omega(\mymatrix{A}) \right] 
    }{
    \omega(\mymatrix{A}) 
    \left[ \mathbb{E}_{12}\, \omega(\mymatrix{A}) \right] 
    } 
  = 
  \frac{ 1 }{ 1 - \alpha_{1}\alpha_{2} } 
  \;  
  \left[
    1 
    - 
    \frac{ \alpha_{1}\alpha_{2} }{  \left( \mathbb{E}_{12}U \right) V }
    \left( \mathbb{E}_{12}\, Q \right) R
  \right]\!.  
\end{equation}
It is easy to see that these equations become (\ref{hie-PP}) if the following 
conditions hold: 
\begin{equation} 
  \left\{
  \begin{array}{l} \displaystyle
  \frac{ \left( \xi_{1} - \xi_{2} \right) \alpha_{k} }
       { \left( \alpha_{1} - \alpha_{2} \right)
         \left( 1 - \alpha_{1}\alpha_{2} \right) }
  \; 
  \frac{ \mathbb{E}_{12}\, U }{ \mathbb{E}_{k}\, U } 
  = 
  \xi_{k} 
  \qquad (k = 1,2),
  \\[6mm] 
  \alpha_{1}\alpha_{2} 
  = 
  - \xi_{1}\xi_{2} \left( \mathbb{E}_{12}\, U \right) V. 
  \end{array}
  \right.
\label{syst-PP}
\end{equation}
The simplest way to satisfy these equations is to take 
\begin{equation}
  \frac{ \mathbb{E}_{\xi} U }{ U } = 
  f(\xi) 
\label{syst-PP-U}
\end{equation}
and 
\begin{equation}
  \alpha(\xi) = \alpha_{*} \xi f(\xi)
\label{syst-PP-A}
\end{equation}
which reduces (\ref{syst-PP}) to 
\begin{equation}
  \alpha_{*}^{2} = - UV = \mbox{constant}
\end{equation}
and 
\begin{equation}
  \left( \xi_{1} - \xi_{2} \right) 
  f\left( \xi_{1} \right)f\left( \xi_{2} \right) 
  = 
  \left[ 
    1 - \alpha_{*}^{2} \xi_{1}\xi_{2} f\left( \xi_{1} \right)f\left( \xi_{2} \right)
  \right] 
  \left[ 
    \xi_{1} f\left( \xi_{1} \right) 
    -  
    \xi_{2} f\left( \xi_{2} \right) 
  \right]. 
\end{equation}
The last equation can be transformed, in $\xi_{2} \to 0$ limit,  into 
``ordinary" one,  
\begin{equation} 
  \alpha_{*}^{2} \xi^{2} f^{2}(\xi)
  + (f_{*} \xi - 1) f(\xi) + 1 = 0
\label{syst-PP-F}
\end{equation}
with an arbitrary constant $f_{*}$. Solution of this quadratic equation that 
satisfies $f(0) = 1$ determines the dependence of $\alpha$ on $\xi$. 

In a similar way one can show that $R$ defined in \eref{def-QR} satisfies the
second of equations \eref{hie-PP}.
Thus, definitions (\ref{def-QR}) together with (\ref{syst-PP-U}), 
(\ref{syst-PP-A}) and (\ref{syst-PP-F}) provide $N$-dark-soliton solutions 
for the functional equations (\ref{hie-PP}) describing the classical AKNS 
hierarchy. Below one can find a more detailed version of these formulae written 
down for the physically relevant case $R = - \overline{Q}$.

\subsection{Solution of the equations of the negative subhierarchy.}

Assuming 
\begin{equation} 
  \overline{\mathbb{E}}_{\eta} \, \mymatrix{A} = 
  \mymatrix{A} \, \mymatrix{H}_{\mymu}^{-1} \mymatrix{H}_{\beta} 
\end{equation} 
where $\beta = \beta(\eta)$ and $\mymu = \beta(0)$ one can obtain from 
(\ref{fay}) with $\xi=\beta$, $\eta=\mymu$, $\zeta=0$ and the shift 
$\mymatrix{A} \to \mymatrix{A} \mymatrix{H}_{\mymu}^{-1}$
\begin{equation}
  ( \beta - \mymu ) 
  \omega(\mymatrix{A}\mysup{r}) 
  \left[ \overline{\mathbb{E}}_{\eta} \, \omega(\mymatrix{B}\mysup{q}) \right]
  = 
  \beta 
  \omega(\mymatrix{A}) 
  \left[ \overline{\mathbb{E}}_{\eta} \, \omega(\mymatrix{B}) \right]
  - 
  \mymu 
  \omega(\mymatrix{B}) 
  \left[ \overline{\mathbb{E}}_{\eta} \, \omega(\mymatrix{A}) \right]. 
\label{fay-NB} 
\end{equation}
In a similar way, Eq. (\ref{fay}) with $\xi=\beta$, $\eta=1/\mymu$, 
$\zeta=0$ and 
$\mymatrix{A} \to \mymatrix{A} \mymatrix{H}_{0}^{-1}$ leads to
\begin{equation}
  \left( 1 - \mymu\beta \right) 
  \omega(\mymatrix{A}) 
  \left[ \overline{\mathbb{E}}_{\eta} \, \omega(\mymatrix{A}) \right]
  = 
  \omega(\mymatrix{A}\mysup{r}) 
  \left[ \overline{\mathbb{E}}_{\eta} \, \omega(\mymatrix{A}\mysup{q}) \right]
  - 
  \mymu\beta 
  \omega(\mymatrix{C}\mysup{r}) 
  \left[ \overline{\mathbb{E}}_{\eta} \, \omega(\mymatrix{B}\mysup{q}) \right]. 
\label{fay-NA} 
\end{equation}
Here the matrices $\mymatrix{A}$, $\mymatrix{B}$ and $\mymatrix{C}$ are the 
ones defined above while
\begin{equation}
  \begin{array}{l}
	\mymatrix{B}\mysup{q} = \mymatrix{A} \mymatrix{H}_{\mymu}, 
	\\[2mm]
	\mymatrix{A}\mysup{q} = \mymatrix{A} \mymatrix{H}_{1/\mymu}^{-1}, 
  \end{array}
\qquad
  \begin{array}{l}
	\mymatrix{C}\mysup{r} = \mymatrix{A} \mymatrix{H}_{\mymu}^{-1}, 
	\\[2mm]
	\mymatrix{A}\mysup{r} = \mymatrix{A} \mymatrix{H}_{1/\mymu}. 
  \end{array}
\end{equation}
In terms of the functions 
\begin{equation}
	q = 
	u \; 
	\frac{ \omega(\mymatrix{B}\mysup{q}) }{ \omega(\mymatrix{A}\mysup{q}) }, 
	\qquad
	r = 
	v \; \frac{ \omega(\mymatrix{C}\mysup{r}) }{ \omega(\mymatrix{A}\mysup{r}) } 
\end{equation}
these equations can be represented in the form  
\begin{equation}
  \omega(\mymatrix{A})  
  \left[ \overline{\mathbb{E}}_{\eta} \, \omega(\mymatrix{A}) \right] 
  \left[ 
    \frac{ \beta }{ \overline{\mathbb{E}}_{\eta}U } 
    \left( \overline{\mathbb{E}}_{\eta} \, Q \right)  
    - 
    \frac{ \mymu }{ U } Q 
  \right] 
  = 
  \frac{ \beta - \mymu }{ \overline{\mathbb{E}}_{\eta}u } \; 
  \omega(\mymatrix{A}\mysup{r}) 
  \left[ \overline{\mathbb{E}}_{\eta} \, \omega(\mymatrix{A}\mysup{q}) \right]
  \; 
  \overline{\mathbb{E}}_{\eta} q 
\end{equation}
and 
\begin{equation}
  ( 1 - \mymu\beta )
  \omega(\mymatrix{A})  
  \left[ \overline{\mathbb{E}}_{\eta} \, \omega(\mymatrix{A}) \right] 
  = 
  \omega(\mymatrix{A}\mysup{r}) 
  \left[ \overline{\mathbb{E}}_{\eta} \, \omega(\mymatrix{A}\mysup{q}) \right]
  \left[ 
    1 
    - 
    \frac{ \mymu\beta }{ \left(\overline{\mathbb{E}}_{\eta}u \right) v } 
    \left( \overline{\mathbb{E}}_{\eta} \, q \right) r 
  \right] 
\end{equation}
which 
leads to (\ref{hie-PN-A}) after imposing the conditions
\begin{equation}
  \left\{
  \begin{array}{l} 
  \mymu\beta = - \left(\overline{\mathbb{E}}_{\eta}u \right) v, 
  \\[2mm]
  \beta U = \mymu \left(\overline{\mathbb{E}}_{\eta}U \right)\!, 
  \\[2mm]
  U ( \beta - \mymu )( 1 - \mymu\beta ) 
  = 
  \eta\mymu \; \overline{\mathbb{E}}_{\eta}u. 
  \end{array} 
  \right. 
\end{equation}
These restrictions can be resolved as follows:
\begin{equation}
	\frac{ \overline{\mathbb{E}}_{\eta} U }{ U } 
  = 
	\frac{ \overline{\mathbb{E}}_{\eta} u }{ u } 
  = 
	g(\eta) 
\end{equation}
and
\begin{equation}
  \beta = \mymu g(\eta), 
  \qquad
  \mymu^{2} = - uv, 
  \qquad
  \frac{u}{U} = \frac{v}{V} = g_{*}, 
\end{equation}
where $g_{*}$ is an arbitrary constant and 
$g(\eta)$ is the solution of the quadratic equation 
\begin{equation}
	\left[ g(\eta) - 1 \right]\left[ 1 - \mymu^{2} g(\eta) \right] 
	= 
	g_{*} \eta g(\eta) 
\label{syst-PN-G}
\end{equation}
satisfying $g(0) = 1$.

This completes settling the problem of finding $N$-dark-soliton solutions of 
the extended AKNS hierarchy because one can show by straightforward algebra 
that the functions $q$ and $r$ presented in this section satisfy Eqs. 
(\ref{eq-neg-qr}) as well.

\subsection{$R = - \overline{Q}$ case.}

This section is devoted to the situation that appears in the physical 
applications of the NLSE (and hence, of the whole AKNS hierarchy): 
\begin{equation}
	R\left( \mathrm{t}, \bar{\mathrm{t}} \right) =  
	- \overline{ Q\left( \mathrm{t}, \bar{\mathrm{t}} \right) },
\end{equation}
where overbar stands for the complex conjugation. In this case the background 
solutions $U$ and $V$ can be represented in the form 
\begin{subequations}
\begin{eqnarray}
	U & = & 
	Q_{*} \exp\left[ i \varphi\left( \mathrm{t}, \bar{\mathrm{t}} \right) \right]\!, 
  \\
	V & = & 
	R_{*} \exp\left[ -i \varphi\left( \mathrm{t}, \bar{\mathrm{t}} \right) \right] 
\end{eqnarray}
\end{subequations}
with the constants $Q_{*}$ and $R_{*}$ being related by
\begin{equation}
	R_{*} = - \overline{Q_{*}}. 
\end{equation}
Similar formulae can be written for $q$ and $r$, 
\begin{subequations}
\begin{eqnarray}
	u & = & 
	q_{*} \exp\left[ i \varphi\left( \mathrm{t}, \bar{\mathrm{t}} \right) \right]\!, 
  \\
	v & = & 
	r_{*} \exp\left[ -i \varphi\left( \mathrm{t}, \bar{\mathrm{t}} \right) \right] 
\end{eqnarray}
\end{subequations}
with 
\begin{equation}
	r_{*} = - \overline{q_{*}}.
\end{equation}
The phase $\varphi$ is determined by the equations 
\begin{equation} 
  \left\{
  \begin{array}{lcl}
  \exp\left[ i \left( \mathbb{E}_{\xi} - 1 \right) \varphi \right] 
  & = & 
  f(\xi), 
  \\[2mm]
  \exp\left[ i \left( \overline{\mathbb{E}}_{\eta} - 1 \right) \varphi \right] 
  & = & 
  g(\eta). 
  \end{array}
  \right.
\end{equation}
Presenting $\varphi$ as
\begin{equation}
  \varphi\left( \mathrm{t}, \bar{\mathrm{t}} \right) =  
  \varphi_{0} + 
  \sum_{k=1}^{\infty} 
  \left( \varphi_{k} t_{k} + \tilde\varphi_{k} \bar{t}_{k}\right) 
\end{equation}
one can obtain for the generating function for the coefficients 
$\varphi_{k}$ and $\tilde\varphi_{k}$ the following expressions:
\begin{subequations}
\begin{eqnarray}
  \sum\limits_{k=1}^{\infty} \varphi_{k} \xi^{k} / k  
  & = & - \ln f(\xi), 
\\ 
  \sum\limits_{k=1}^{\infty} \tilde\varphi_{k} \eta^{k} / k 
  & = & - \ln g(\eta) 
\end{eqnarray}
\end{subequations}
or, after applying the $\xi d/d\xi$ and $\eta d/d\eta$ operators and using 
Eqs. \eref{syst-PP-F} and \eref{syst-PN-G} for $f(\xi)$ and $g(\eta)$,
\begin{subequations}
\begin{eqnarray}
  \sum\limits_{k=1}^{\infty} \varphi_{k} \xi^{k} 
  & = & \displaystyle 
  1 - \frac{ f(\xi) }{ 1 - \left| Q_{*} \right|^{2} \xi^{2} f^{2}(\xi) }, 
  \\[4mm] 
  \sum\limits_{k=1}^{\infty} \tilde\varphi_{k} \eta^{k} 
  & = & 
  \frac{ 
    \left[ 1 - g(\eta) \right] 
    \left[ 1 - \left| q_{*} \right|^{2} g(\eta) \right] 
  }{ 
    1 - \left| q_{*} \right|^{2} g^{2}(\eta) 
  } 
\end{eqnarray}
\end{subequations}
with arbitrary real $\varphi_{0}$.
Then, it is easy to check that to ensure the necessary properties of the 
matrices $\mymatrix{A}$ one has to choose
\begin{equation}
  \mymatrix{L} = \mbox{diag} ( e^{i\theta_{n}} )_{n = 1, ..., N}. 
\end{equation}
In this case the matrices describing the $t$- and 
$\bar{t}$-evolution are unitary, 
\begin{equation}
  \mymatrix{H}_{0}^{-1} \mymatrix{H}_{\alpha(\xi)} =  
  \mbox{diag} 
  \left( \; e^{ i\phi_{n}(\xi) } \; \right)_{n = 1, ..., N} 
\end{equation}
and 
\begin{equation}
  \mymatrix{H}_{\mymu}^{-1} \mymatrix{H}_{\beta(\eta)} 
  =  
  \mbox{diag} 
  \left( \; e^{ i \left[ \psi_{n}(\eta) - \psi_{n}(0) \right] } \; \right)_{n = 1, ..., N} 
\end{equation}
with 
\begin{subequations}
\begin{eqnarray}
  \phi_{n}(\xi) & = & 
  2\arg\left( \; 
    1 - \left| Q_{*} \right| \xi f(\xi) e^{ -i\theta_{n} } 
  \; \right)\!,
  \\[2mm]
  \psi_{n}(\xi) & = & 
  2\arg\left( \; 
    1 - \left| q_{*} \right| g(\eta) e^{ -i\theta_{n} } 
  \; \right)\!. 
\end{eqnarray} 
\end{subequations}
Now one can establish the dependence of the matrix $\mymatrix{A}$ on the variables 
$t_{k}$ and $\bar{t}_{k}$:
\begin{equation}
  \mymatrix{A}\left( \mathrm{t}, \bar{\mathrm{t}} \right) = 
  \mymatrix{A}_{*} \, 
  \mbox{diag}\left( 
    e^{ \nu_{n}\left( \mathrm{t}, \bar{\mathrm{t}} \right) } 
  \right)_{n = 1, ..., N} ,  
\end{equation}
where $\mymatrix{A}_{*}$ is a constant matrix and 
\begin{equation}
  \nu_{n}\left( \mathrm{t}, \bar{\mathrm{t}} \right) = 
  \nu_{n0} + 
  \sum\limits_{k=1}^{\infty} 
  \left( \nu_{nk} t_{k} + \tilde\nu_{nk} \bar{t}_{k} \right) 
\end{equation}
with arbitrary real $\nu_{n0}$ and 
\begin{subequations}
\begin{eqnarray}
  \sum\limits_{k=1}^{\infty} \nu_{nk} \xi^{k} / k  
  & = & \phi_{n}(\xi), 
  \\
  \sum\limits_{k=1}^{\infty} \tilde\nu_{nk} \eta^{k} / k  
  & = & \psi_{n}(\eta) - \psi_{n}(0) 
\end{eqnarray}
\end{subequations}
or 
\begin{subequations}
\begin{eqnarray}
  \sum\limits_{k=1}^{\infty} \nu_{nk} \xi^{k} 
  & = & 
  \xi \phi_{n}'(\xi), 
  \\
  \sum\limits_{k=1}^{\infty} \tilde\nu_{nk} \eta^{k} 
  & = & 
  \eta \psi_{n}'(\eta). 
\end{eqnarray}
\end{subequations}
Upon noting that the determinants (\ref{omega-def}) are invariant under 
transformations $\mymatrix{A} \to \mymatrix{M}^{-1} \mymatrix{A} \mymatrix{M}$ 
one can eliminate, without loss of generality, the constants 
$\ell_{m}$ ($\ell_{m} \to 1$) by redefining the functions 
$a_{n}\left( \mathrm{t}, \bar{\mathrm{t}} \right)$
($a_{n}\left( \mathrm{t}, \bar{\mathrm{t}} \right) \to 
a_{n}\left( \mathrm{t}, \bar{\mathrm{t}} \right) \ell_{n}$)
thus arriving at the final expressions for the N-dark-soliton solutions of the 
extended AKNS hierarchy: 
\begin{eqnarray}
  Q\left( \mathrm{t}, \bar{\mathrm{t}} \right) 
  & = & 
  \left|Q_{*}\right| 
    e^{i\varphi\left( \mathrm{t}, \bar{\mathrm{t}} \right)} 
    \frac{\Delta_{1}\left( \mathrm{t}, \bar{\mathrm{t}} \right)}
         {\Delta_{0}\left( \mathrm{t}, \bar{\mathrm{t}} \right)},
\\
  R\left( \mathrm{t}, \bar{\mathrm{t}} \right) & = & 
  - \left|Q_{*}\right| 
    e^{-i\varphi\left( \mathrm{t}, \bar{\mathrm{t}} \right)} 
    \frac{\Delta_{-1}\left( \mathrm{t}, \bar{\mathrm{t}} \right)}
         {\Delta_{0}\left( \mathrm{t}, \bar{\mathrm{t}} \right)}.
\end{eqnarray}
Here, the determinants $\Delta_{\epsilon}$ ($\epsilon = 0, \pm 1$) 
are given by 
\begin{equation}
  \Delta_{\epsilon}\left( \mathrm{t}, \bar{\mathrm{t}} \right) = 
  \det\left| 
    \delta_{mn} + 
    \frac{ \exp\left( \nu_{n}\left( \mathrm{t}, \bar{\mathrm{t}} \right) 
           + 2i\epsilon\theta_{n} \right) }
         { \sin\left( \frac{ \theta_{m} + \theta_{n} }{ 2 } \right) } 
  \right|_{m,n =1, ..., N}
\end{equation}
while the phase of $Q_{*}$ and the real constants 
$a_{n}( \mathrm{0}, \bar{\mathrm{0}} )$ are 
respectively absorbed into 
$\varphi_{0}$ and $\nu_{n0}$.

\appendix

\section{Proof of \eref{fay}. \label{app-Fay}}

In this appendix we present some identities for the matrices defined in Sec. 
\ref{sec-dark} which provide a proof of the Fay's identity \eref{fay}. 
Consider the matrix $\mymatrix{A}$ satisfying 
\begin{equation}
  \mymatrix{L} \mymatrix{A} 
  - 
  \mymatrix{A} \mymatrix{M} 
  = 
  | \,\ell\, \rangle \langle a |
\label{app-A}
\end{equation}
with arbitrary diagonal matrices $\mymatrix{L}$ and $\mymatrix{M}$ together 
with the matrices $\mymatrix{H}_{\zeta}$ defined by 
\begin{equation}
  \mymatrix{H}_{\zeta} 
  = 
  \left( \zeta - \mymatrix{L} \right)
  \left( \zeta - \mymatrix{M} \right)^{-1}, 
\end{equation}
where $\left( \zeta - \mymatrix{L} \right)$ stands for 
$\left( \zeta \mymatrix{1} - \mymatrix{L} \right)$ etc.
It follows from \eref{app-A} that 
\begin{equation}
  (\zeta - \mymatrix{M})
  \left( \mymatrix{1} + \mymatrix{H}_{\zeta} \mymatrix{A} \right) 
  (\zeta - \mymatrix{M})^{-1} = 
  \mymatrix{1} 
  + \mymatrix{A} 
  - | \,\ell\, \rangle \langle b_{\zeta} | 
\label{app-HA}
\end{equation}
with 
\begin{equation}
  \langle b_{\zeta} | 
  =  
  \langle a | (\zeta - \mymatrix{M})^{-1} 
\end{equation}
which leads to 
\begin{equation}
  \det\left| \mymatrix{1} + \mymatrix{A} \mymatrix{H}_{\zeta} \right| 
  = 
  \det\left| \mymatrix{1} + \mymatrix{A} \right| 
  \cdot
  \det\left| \mymatrix{1} - | \,e\, \rangle \langle b_{\zeta}| \, \right| 
\end{equation}
with
\begin{equation}
  | \,e\, \rangle = 
  \left( \mymatrix{1} + \mymatrix{A} \right)^{-1} | \,\ell\, \rangle 
\end{equation}
and hence to 
\begin{equation}
  \frac{ \omega_{\zeta} }{ \omega } 
  = 
  1 - \langle b_{\zeta} | \,e\, \rangle.
\label{app-Z}
\end{equation}
A little bit more cumbersome calculations lead to the following ``two-point" 
analogue of \eref{app-HA}: 
\begin{eqnarray}
&&
  (\xi - \mymatrix{M})(\eta - \mymatrix{M}) 
  \left( 
    \mymatrix{1} 
    + \mymatrix{H}_{\xi} \mymatrix{H}_{\eta} \mymatrix{A} 
  \right) 
  (\xi - \mymatrix{M})^{-1} (\eta - \mymatrix{M})^{-1} 
\\&&\qquad 
  = 
  \mymatrix{1} + \mymatrix{A} 
  + \frac{ \xi - \mymatrix{L} }{ \eta - \xi } 
    | \,\ell\, \rangle \langle b_{\eta} |
  + \frac{ \eta - \mymatrix{L} }{ \xi - \eta } 
    | \,\ell\, \rangle \langle b_{\xi} | 
\end{eqnarray}
and 
\begin{equation}
  \frac{ \omega_{\xi\eta} }{ \omega } =
  \det \biggl| 
  \mymatrix{1} 
  +  | \,u_{1} \, \rangle \langle v_{1} | 
  +  | \,u_{2} \, \rangle \langle v_{2} | 
  \biggr|. 
\label{app-XY}
\end{equation}
Here, the rows $\langle v_{1,2} |$ and the columns $| \,u_{1,2} \rangle$ are 
defined by 
\begin{equation}
  \langle v_{1} | = \langle b_{\eta} |, 
  \qquad
  \langle v_{2} | = \langle b_{\xi} | 
\end{equation}
and 
\begin{equation}
  | \,u_{1} \rangle  = 
  \displaystyle \frac{ 1 }{ \eta - \xi } 
  | \, c_{\xi} \, \rangle, 
  \qquad
  | \,u_{2} \rangle = 
  \displaystyle \phantom{-} \frac{ 1 }{ \xi - \eta } 
  | \, c_{\eta} \, \rangle 
\end{equation}
with 
\begin{equation}
  | \, c_{\zeta} \rangle  = 
  \left( \mymatrix{1} + \mymatrix{A} \right)^{-1}  
  \left(\zeta - \mymatrix{L} \right) 
  | \,\ell\, \rangle. 
\end{equation}
Rewriting the determinant in the right-hand side of \eref{app-XY} as 
\begin{equation}
  \frac{ \omega_{\xi\eta} }{ \omega } 
  = 
  \left| 
    \begin{array}{cc}
    1 + \langle v_{1} | \,u_{1} \rangle & 
    \langle v_{1} | \,u_{2} \rangle \\
    \langle v_{2} | \,u_{1} \rangle & 
    1 + \langle v_{2} | \,u_{2} \rangle 
    \end{array}
    \right| 
\end{equation}
and calculating the scalar products, 
\begin{eqnarray}
  \langle v_{2} | \,u_{1} \rangle & = & 
  \frac{ \varphi_{\xi} }{ \eta - \xi }, 
\\
  \langle v_{1} | \,u_{2} \rangle & = & 
  \frac{ \varphi_{\eta} }{ \xi - \eta } 
\end{eqnarray}
with 
\begin{equation}
  \varphi_{\zeta} 
  = 
  \langle b_{\zeta} | c_{\zeta} \rangle 
\end{equation}
and 
\begin{eqnarray}
  1 + \langle v_{1} | \,u_{1} \rangle & = & 
  \frac{ \omega_{\eta} }{ \omega } + \frac{ \varphi_{\eta} }{ \eta - \xi }, 
  \\ 
  1 + \langle v_{2} | \,u_{2} \rangle & = & \displaystyle 
  \frac{ \omega_{\xi} }{ \omega } + \frac{ \varphi_{\xi} }{ \xi - \eta } 
\end{eqnarray}
(here Eq. \eref{app-Z} was used) one arrives at 
\begin{equation}
  \frac{ \omega \, \omega_{\xi\eta} }{ \omega_{\xi} \, \omega_{\eta} } 
  = 
  1 + 
  \frac{ \omega  }{ \xi - \eta} 
  \left[ 
    \frac{ \varphi_{\xi} }{ \omega_{\xi} } 
    - 
    \frac{ \varphi_{\eta} }{ \omega_{\eta} } 
  \right] 
\end{equation}
which leads to ``separation of variables",
\begin{equation}
  \left( \xi - \eta \right) 
  \frac{ \omega \, \omega_{\xi\eta} }{ \omega_{\xi} \, \omega_{\eta} } 
  = 
  \Omega_{\xi} - \Omega_{\eta} 
\label{app-main}
\end{equation}
where 
\begin{equation}
  \Omega_{\zeta} = \zeta + \frac{ \omega \varphi_{\zeta} }{ \omega_{\zeta} }. 
\end{equation}
Upon adding three copies of \eref{app-main} for 
$(\xi,\eta)$, $(\eta,\zeta)$ and $(\zeta,\xi)$ one can obtain the identity 
\eref{fay} that we want to prove.


\end{document}